\documentclass[12pt]{article}

\usepackage{amsmath}
\usepackage{amssymb}
\usepackage{graphicx}

\def\PL{ Phys. Lett. }
\def\PR{ Phys. Rev. }
\def\PRL{ Phys. Rev. Lett. }
\def\APJ{ Astroph.~J.~}
\def\AJ{ Astron.~J.~}

\begin{document}

\title{Stringy Dark Energy Model with Cold Dark Matter}

\author{
I.~Ya.~Aref'eva\footnote{\texttt{arefeva@mi.ras.ru}, Steklov Mathematical
Institute, Russian Academy of Sciences},
A.~S.~Koshelev\footnote{\texttt{koshelev@mi.ras.ru}, Steklov Mathematical
Institute, Russian Academy of Sciences}
\\and\\
S.~Yu.~Vernov\footnote{\texttt{svernov@theory.sinp.msu.ru}, Skobeltsyn
Institute of Nuclear Physics, Moscow State University}}

\date{}

\maketitle

\begin{abstract}
Cosmological consequences of adding the Cold Dark Matter (CDM) to the  exactly
solvable stringy Dark Energy (DE) model are investigated. The model is
motivated by the consideration of our Universe as a slowly decaying D3-brane.
The  decay of this D-brane is described in the String Field Theory
framework. Stability conditions of the exact solution with respect to small
fluctuations of the initial value of the CDM energy density are found.
 Solutions with large initial value of the CDM
energy density attracted by the exact solution without CDM are constructed
numerically. In contrast to the $\Lambda$CDM model the Hubble parameter in the
model is not a monotonic function of time. For specific initial data the DE
state parameter $w_{DE}$ is also not monotonic function of time. For these
cases there are two separate regions of time where $w_{DE}$ being less than $-1$ is close to
$-1$.
\end{abstract}

\thispagestyle{empty}


\newpage

\section{Introduction}

Nowadays strings and D-branes found cosmological applications related with the
cosmological acceleration \cite{Sen}-\cite{IA1}.
The combined analysis of the type Ia supernovae, galaxy clusters measurements
and WMAP data provides an evidence for the accelerated cosmic
expansion~\cite{Riess,Perlm,Spergel}. The cosmological acceleration strongly
indicates that the present day Universe is dominated by smoothly distributed
slowly varying Dark Energy (DE) component~(see \cite{Sachni} for reviews), for
which the state parameter $w_{DE}$ is negative\footnote{Here $w_{DE}$ is usual notation
for the pressure to energy ratio.}. Contemporary experiments give strong
support that currently the state parameter $w_{DE}$ is close to $-1$, $w_{DE}=-1\pm 0.1$
\cite{Spergel,Tonry,Tegmark,Riess2,Seljak}.

From the theoretical point of view the specified domain of $w$ covers three
essentially different cases: $w>-1,~w=-1$ and $w<-1$ (see~\cite{AKV}, and
references therein). The most exciting possibility would be the case $w < -1$
corresponding to the so called phantom dominated Universe. In phenomenological
models describing this case the weak energy conditions $\varrho>0,
\varrho+p>0$ are violated and there are problems with stability at classical
and quantum levels~\cite{phantom}. Thus, a phantom becomes a great challenge
for the theory while its presence according to the supernovae data is not excluded.

A possible way to evade the stability problem for a phantom model is to yield
the phantom as an effective model of a more fundamental theory which has no
such problems at all. It has been shown in~\cite{IA1} that such a model does
appear in the string theory framework. This DE model assumes that our Universe
is a slowly decaying D3-brane which dynamics is described by the tachyonic
mode of the string field theory (SFT). The notable feature of the SFT
description of the tachyon dynamics is a non-local polynomial interaction
\cite{Witten-SFT}-\cite{SFT-review}. It turns out the string tachyon behavior
is effectively described by a scalar field with a negative kinetic term
(phantom) however due to the string theory origin the model is stable at large
times.

In \cite{AKV} we have found an exactly solvable Stringy DE model in the
Friedmann Universe. This model is a modified version of the effective SFT
model \cite{IA1} and is inspired by SuperSFT calculations \cite{ABKM}. First
level calculations in the SFT give fourth order polynomial interaction. Higher
levels increase a power of the interaction. Exactly solvable model has a
particular six order polynomial interaction potential. However, small
fluctuations of coefficients in that potential do not change the solution
qualitatively and one can say that the model \cite{AKV} represents the
behavior  of nonBPS D3 brane in the the Friedmann Universe rather well. It is
interesting to investigate the dynamics of the model in the presence of the
Dark Matter. This is a subject of the present paper.

It turns out from the observational data that DE forms about 73\% and the
Dark Matter forms about 23\% of our Universe. Thus because of a significance
of the Dark Matter component in the Universe in the present paper we
investigate an interaction of the phantom matter considered in \cite{AKV} with
the CDM. It seems impossible to find exact solutions in the presence of the
CDM, except the case when  the DE state parameter is  a constant \cite{S-St},
so we use numeric methods to analyze the behavior of the phantom field and
cosmological parameters in our model.


\section{Exactly solvable Phantom Model}

We start by recalling the main facts related to the model considered in
\cite{AKV}. This is a model of Einstein gravity interacting with a single
\textsl{phantom} scalar field in the spatially flat Friedmann Universe. Since
the phantom field comes from the string field theory the string mass $M_s$ and
a dimensionless open string coupling constant $g_o$ emerges. The action is
\begin{equation}
 S=\int d^4x \sqrt{-g}\left(\frac{M_P^2}{2M_s^2}R+\frac1{g_o^2}\left(
 +\frac{1}{2}g^{\mu\nu}\partial_{\mu}\phi\partial_{\nu}\phi
-V(\phi)\right)\right), \label{action}
\end{equation}
where $M_P$ is the reduced Planck mass, $g_{\mu\nu}$ is a spatially flat
Friedmann metric
\begin{equation*}
 ds^2={}-dt^2+a^2(t)(dx_1^2+dx_2^2+dx_3^2).
\end{equation*}
and coordinates $(t,x_i)$ and field $\phi$ are dimensionless. Hereafter we use
the dimensionless parameter $m_p$ for short:
\begin{equation}
m_p^2=\frac{g_o^2M_P^2}{M_s^2}. \label{m_p}
\end{equation}
If the scalar field depends only on time, i.e. $\phi=\phi(t)$, then
independent equations of motion are
\begin{equation}
\begin{split}
3H^2&=\frac{1}{m_p^2}\:\varrho_{DE},\quad
\varrho_{DE}=-\frac12\dot\phi^2+V(\phi),
\\3H^2+2\dot H&={}-\frac{1}{m_p^2}\:p_{DE},\quad p_{DE}=-\frac12\dot\phi^2-V(\phi),
\end{split}
\label{eomprho}
\end{equation}
Here dot denotes the time derivative, $H\equiv \dot a(t)/a(t)$, $\varrho_{DE}$
and $p_{DE}$ are energy and pressure densities of the DE respectively. One can
recast the system (\ref{eomprho}) to the following form
\begin{equation}
\begin{split}
  \dot H&=\frac{1}{2m_p^2}\dot\phi^2,\\
  3H^2&=\frac{1}{m_p^2}\left(-\frac{1}{2}\dot\phi^2+V(\phi)\right).
\end{split}
\label{eom}
\end{equation}
Besides of this there is an equation of motion for the field $\phi$ which is
in fact a consequence of system (\ref{eomprho}).

Following the superpotential method \cite{DeWolfe} (see also
\cite{padmanabhanfirst}) we assume that $H(t)$ is a function (named as
superpotential) of $\phi(t)$:
\begin{equation}
 H(t)=W(\phi(t)).
\end{equation}

This still does not give a systematic way to find general solutions to the
system (\ref{eom}) but allows one to construct $W(\phi)$ and $V(\phi)$ for a
known function $\phi(t)$. We take for $\phi(t)$
\begin{equation}
\phi(t)=A\tanh(\omega t). \label{kinkphi}
\end{equation}
This function is known to describe effectively the late time behavior of the
tachyon in the 4-dimensional flat case \cite{AJK,yar}. The function $\phi(t)$ satisfies the
following equation
\begin{equation*}
\dot\phi=\omega\left(A-\frac{1}{A}\phi^2\right).
\end{equation*}
Hence, we obtain
\begin{equation}
 W=\frac{\omega}{2m_p^2}\left(A\phi-\frac{1}{3A}{\phi}^{3}
\right),\label{ourW}
\end{equation}
and corresponding potential
\begin{equation}
V(\phi)= \frac{\omega^2}{2A^2}\left(A^2-\phi^2\right)^2
+\frac{\omega^2\phi^2}{12A^2m_p^2}\left( 3\,A^2-\phi^2 \right)^2. \label{ourV}
\end{equation}
We have omitted an integration constant in (\ref{ourW}) to yield an even
potential~(\ref{ourV}). It is typical that to keep the form of solutions to
the scalar field equation in the presence of Friedmann metric one has to
modify the potential adding a term proportional to the inverse of the reduced
Planck mass $M_P^2$ \cite{AKV,AJ}.

The described solution leads to a number of cosmological consequences. The
Hubble parameter
\begin{equation}
H=\frac{\omega A^2}{2m_p^2}\tanh(\omega t)\left(1-\frac{1}{3}(\tanh(\omega
t))^2\right)
\end{equation}
goes asymptotically to $\omega A^2/(3m_p^2)$ when $t$ goes to infinity. Once
$H(t)$ is known one readily obtains the scale factor
\begin{equation}
a(t)=a_0(\cosh(\omega t))^{\frac{A^2}{3m_p^2}}\exp\left({\frac{
A^2(\cosh(\omega t)^2-1)} {12m_p^2\cosh(\omega t)^2}}\right), \label{func_a}
\end{equation}
where $a_0$ is an arbitrary constant, and the deceleration parameter
\begin{equation}
q(t)=-\:\frac{\ddot{a}a}{\dot{a}^2}={}-1-{\frac {18m_p^2 \left(
\cosh(\omega\,t)\right)^{2}}{A^2\left(\left( \cosh(\omega\,t)
 \right) ^{2}-1 \right)  \left( 2\, \left( \cosh(\omega\,t)\right)^2
 +1 \right)^2}}. \label{deceleration}
\end{equation}

It follows from formula (\ref{deceleration}) that the Universe in this
scenario is accelerating.

The expression for the state parameter is the following
\begin{equation}
w_{DE}(\phi)=\frac {p_{DE}(\phi)}{\varrho_{DE}(\phi)}=-1
-12m_p^2\frac{(A^2-\phi^2)^2}{\phi^2(3A^2-\phi^2)^2}.
\end{equation}
Point $\phi=A$ corresponds to an infinite future and therefore $w_{DE}\to -1$
as $t\to\infty$.

\begin{figure}[h]
\centering
\includegraphics[width=40mm]{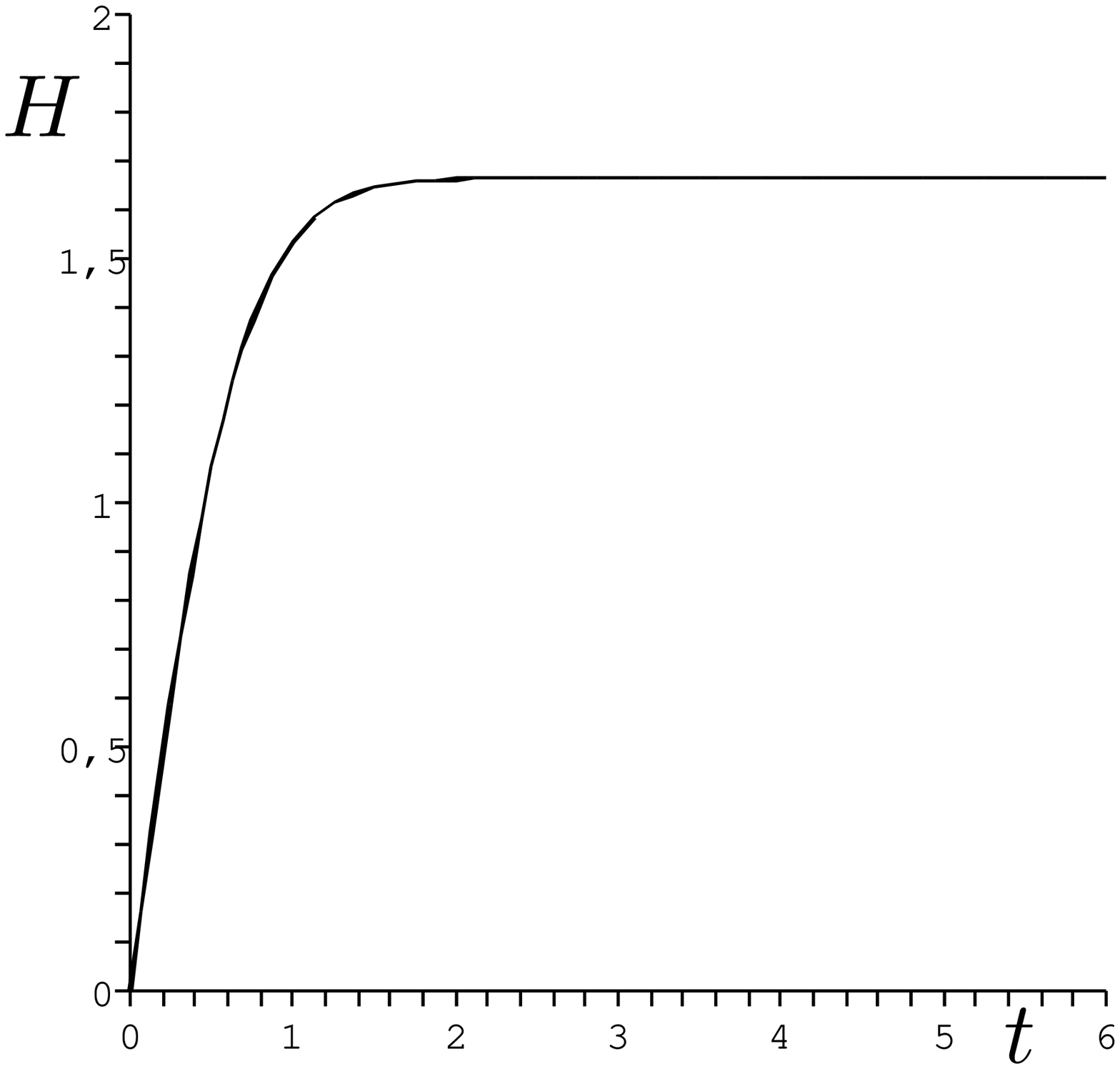}
\includegraphics[width=40mm]{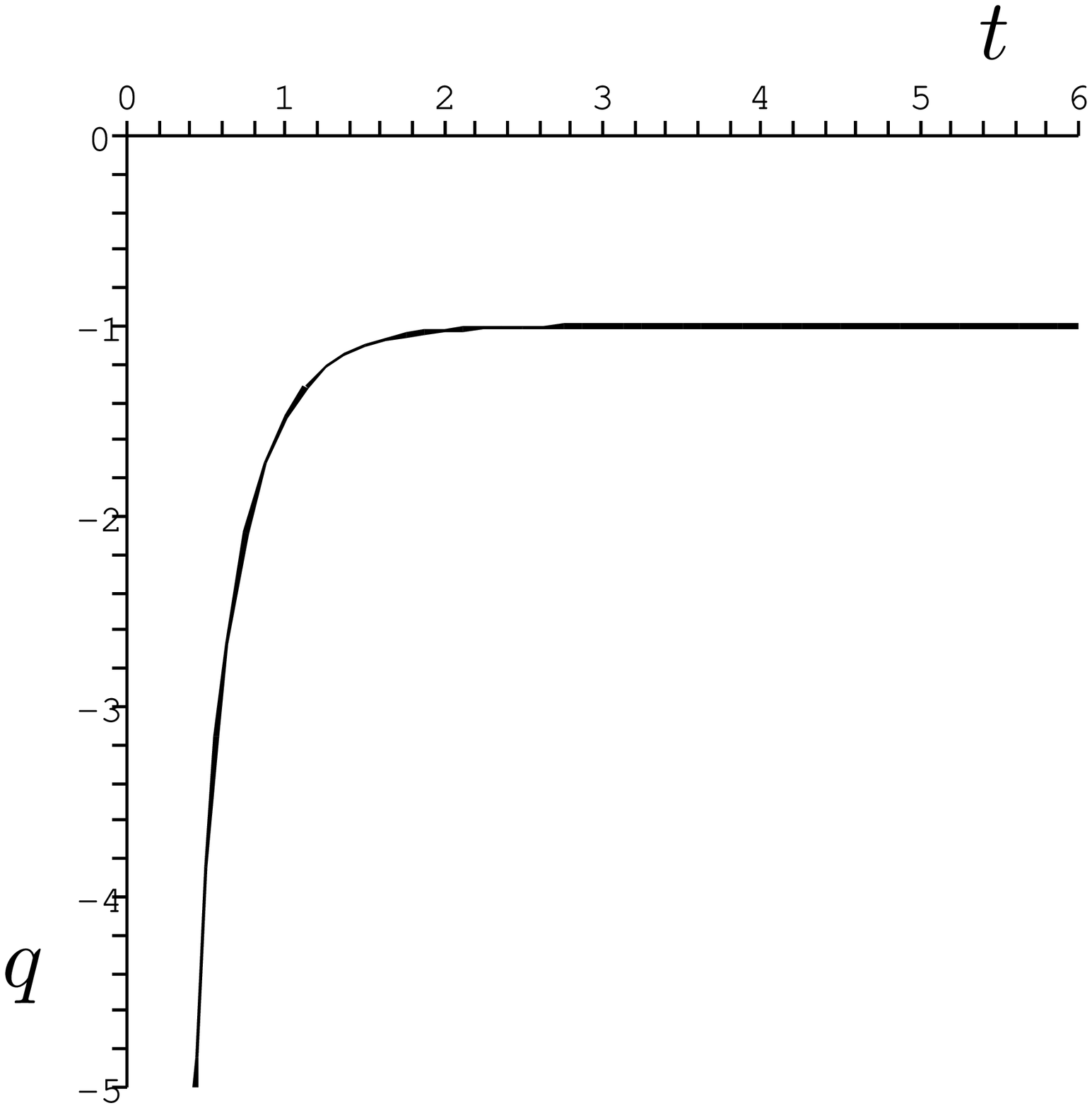}
\includegraphics[width=40mm]{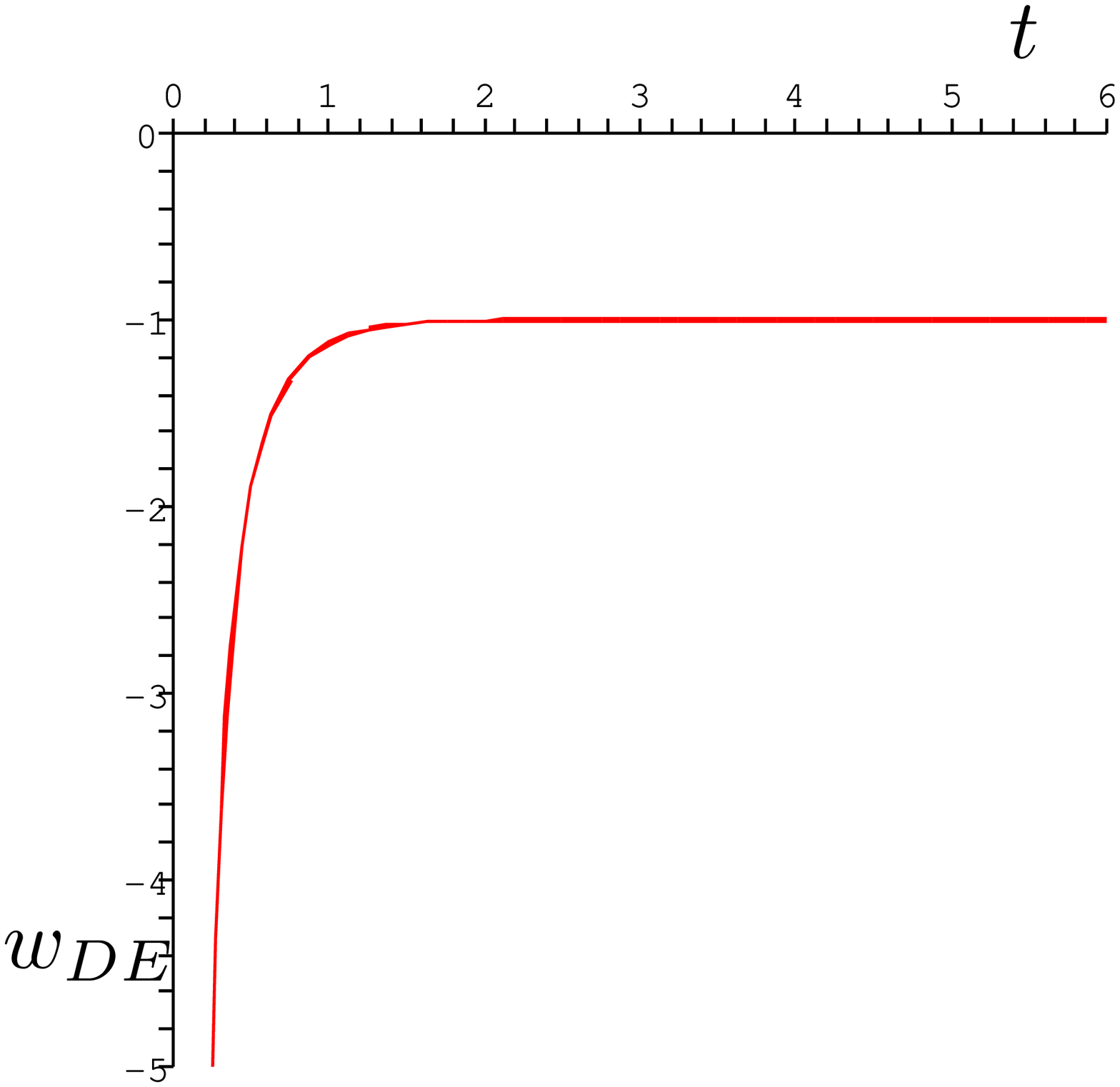}
\caption{The time evoution of the Hubble parameter $H(t)$ (left), the deceleration parameter $q(t)$ (middle)
and the state parameter $w_{DE}(t)$ (right) in the exactly solvable model for $m_p^2=0.2$.}
\label{plots}
\end{figure}

Plots for the Hubble, deceleration and state parameters are drawn in Fig.\ref{plots}
(Hereafter we assume $A=\omega=1$ for all plots).

Thus, we conclude by noting that in our model the phantom field provides the
DE dominated accelerating Universe.


\section{Interaction with Cold Dark Matter}

\subsection{The Model}

Now we are going to couple in a minimal way a pressureless matter of energy
density $\varrho_{M}$ (the CDM) to our model such that the Friedmann equations
get an extra term
\begin{eqnarray}
  3H^2&=&\frac{1}{m_p^2}\left(-\frac{1}{2}\dot\phi^2+V(\phi)+
\varrho_M(a)\right),\label{eom-cdm}
\\ \dot H&=&\frac{1}{2m_p^2}\left(\dot\phi^2-\varrho_M(a)\right).\label{eom-cdm2}
\end{eqnarray}
and the equation describing the evolution of scalar field has previous form
\begin{equation}
\ddot{\phi}+3H\dot\phi-V_{\phi}^{\prime}=0. \label{eomphi2order-nm}
\end{equation}

From (\ref{eom-cdm})--(\ref{eomphi2order-nm}) we obtain the conservation of
the energy density for the CDM:
\begin{equation}
\dot\varrho_M+3H\varrho_M=0,
\end{equation}
that after integration gives
\begin{eqnarray}
\varrho_M=\varrho_{M,0}e^{-3\int\limits^t H(\tau) d
\tau}=\varrho_{M,0}\left(\frac{a}{a_0}\right)^{-3}, \label{rho}
\end{eqnarray}
where constants $\varrho_{M,0}$ and $a_0$ are initial values of $\varrho_M$
and $a$ correspondingly.
From (\ref{rho}) we obtain equation (\ref{eom-cdm}) in the
following form:
\begin{equation}
3H^2=\frac{1}{m_p^2}\left(V(\phi)-\frac{~\dot\phi^2}{2}+\varrho_{M,0}\left(\frac{a_0}{a}
\right)^3\right). \label{eom-cdm-expl}
\end{equation}

Following the lines of \cite{AKV} we address to our analysis the questions of
cosmological evolution and stability.

The straightforward way to study a stability of solutions to the system of
equations (\ref{eom-cdm})--(\ref{eomphi2order-nm}) is to exclude $\varrho_M$
from (\ref{eom-cdm}),(\ref{eom-cdm2}) and obtain the following system:
\begin{eqnarray}
\ddot{\phi}&+&3H\dot\phi-V_{\phi}^{\prime}=0, \label{phi2order}
\\  2\dot H+3H^2&=&\frac{1}{m_p^2}\left(\frac{1}{2}\dot\phi^2+V(\phi)\right).
\label{Hequ}
\end{eqnarray}
Depending on the initial values of $H$, $\phi$ and $\dot\phi$, which have been
considered as independent, this system describes our model either with or
without the CDM.  In particular, the initial values: $H_0=0$, $\phi_0=0$ and
$\dot\phi_0=A\omega$ correspond to the exact solution (\ref{kinkphi}).

Either one can perform calculations using nonautonomous system of equations
\cite{Mukh-rep} which can be obtained from equations (\ref{eomphi2order-nm})
and (\ref{eom-cdm-expl}) in the following form
\begin{eqnarray}
\frac{d \phi}{d n}&=&\frac{1}{H(\phi,\psi,n)}~\psi, \label{fo-1}
\\  \frac{d \psi}{d n}&=&{}-3\psi+\frac{1}{H(\phi,\psi,n)}~\frac{d V(\phi)}{d \phi},
\label{fo-2}
\end{eqnarray}
where
\begin{equation*}
H(\phi,\psi,n)=\frac{1}{\sqrt{3}M_p}\sqrt{-\frac{1}{2}\psi^2+V(\phi)+\varrho_{M,0}e^{-3n}},
~\dot{\phi}=\psi~\text{and}~n=\ln (a/a_0).
\end{equation*}

\subsection{Stability analysis for small fluctuations.}

In~\cite{AKV} we have analyzed the system
(\ref{eom-cdm})--(\ref{eomphi2order-nm}) without the CDM under condition
$A=\omega=1$ and found that the exact solution $\phi=\tanh(t)$ is stable with
respect to small fluctuations of the initial conditions if and only if
$m_p^2\leqslant 1/2$.

Let us consider the behavior of the solution of system
(\ref{phi2order})--(\ref{Hequ}) in the neighborhood of the exact solution
\begin{equation*}
\begin{split}
\phi_0(t)&=\tanh(t),\\
\dot\phi_0(t)&\equiv\psi_0(t)=1-\tanh(t)^2,\\
H_0(t)&=\frac{1}{2m_p^2}\tanh(t)\left(1-\frac{1}{3}\tanh(t)^2\right).
\end{split}
\end{equation*}
Substituting
\begin{equation}
  H(t)=H_0(t)+\varepsilon H_1(t), \quad \phi(t)=\phi_0(t)+\varepsilon\phi_1(t)
  \quad \mbox{and} \quad
  \dot\phi(t)\equiv\psi(t)=\psi_0(t)+\varepsilon\psi_1(t),
\end{equation}
in (\ref{phi2order}) and (\ref{Hequ}) we obtain in the first order of
$\varepsilon$ the following equations:
\begin{equation}
\begin{array}{@{}rcl@{}}
\displaystyle \dot \phi_1&\displaystyle =&\displaystyle \psi_1,\\[3.7mm]
\displaystyle \dot\psi_1 &\displaystyle=&\displaystyle
\frac{1}{m_p^2}\left(2(2m_p^2-1)-\frac{6m_p^2-1}{\cosh(t)^2}\right)
\phi_1-{}\\[2.7mm]&\displaystyle -&\displaystyle \frac{(2\cosh(t)^2+1)\tanh(t)}
{2m_p^2\cosh(t)^2}\psi_1-\frac{3}{\cosh(t)^2}H_1,\\[3.7mm]
\displaystyle \dot H_1&\displaystyle =&\displaystyle \frac{\tanh(t)}{4m_p^2\cosh(t)^4}\displaystyle
\left(1+(2-4m_p^2)\cosh(t)^2\right)\phi_1+{}\\[2.7mm]
&\displaystyle +&\displaystyle \frac{1}{2\cosh(t)^2}\psi_1 -\frac{(1+2\cosh(t)^2)\tanh(t)}{2\cosh(t)^2}H_1.\\
\end{array}
\label{equeps}
\end{equation}

System~(\ref{equeps}) has been solved with the help of the computer algebra
system Maple. The exact dependence $\phi_1(t)$, $\psi_1(t)$ and $H(t)$ are too
cumbersome to be presented here. The main result is that for $m_p^2\leqslant
1/2$ functions $\phi_1(t)$, $\psi_1(t)$ and $H_1(t)$ are bounded functions and
our exact solution is stable.

Note that numerical calculations show that if $m_p^2\leqslant 1/2$ then even
for large initial values of the CDM energy density numerical solutions tend to
the exact solution as $t$ tends to infinity.

\subsection{Numeric solutions. Time dependence.}

At this point we pass to numeric methods because it seems impossible to find
exact solutions in the presence of the CDM.  To analyze the cosmological
evolution it is instructive to plot phase curves for the scalar field as well
as evolution of the state parameter $w_{DE}$ for the scalar matter. In
addition we find numerically a ratio of the energy densities for the CDM and
the DE. Experimental bounds for this ratio is known and estimated to be near
$1/3$ so we can find the time point we live and a corresponding value of
$w_{DE}$ in our approach.

\begin{figure}[h]
\centering
\includegraphics[width=40mm]{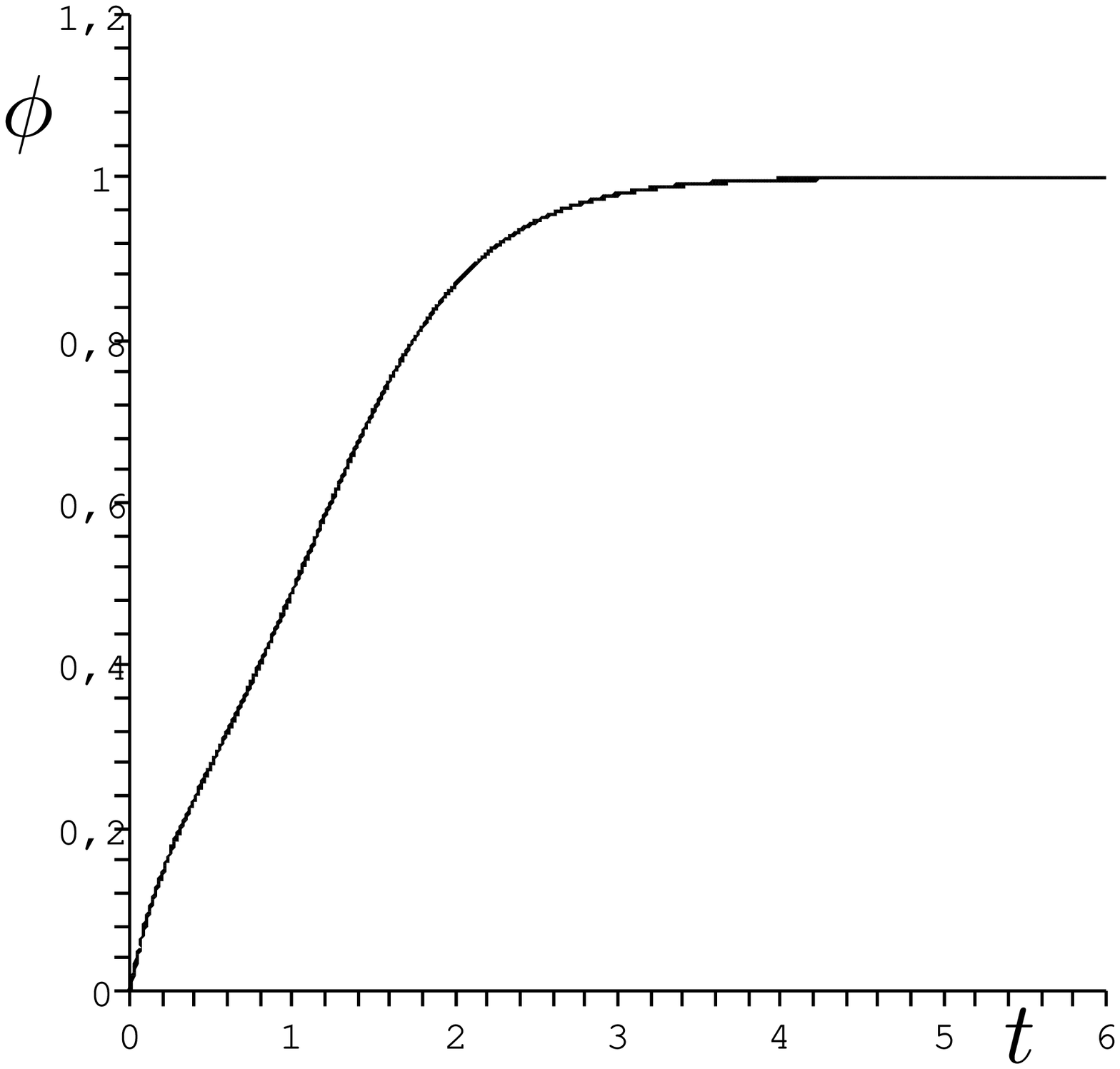} { }
\includegraphics[width=40mm]{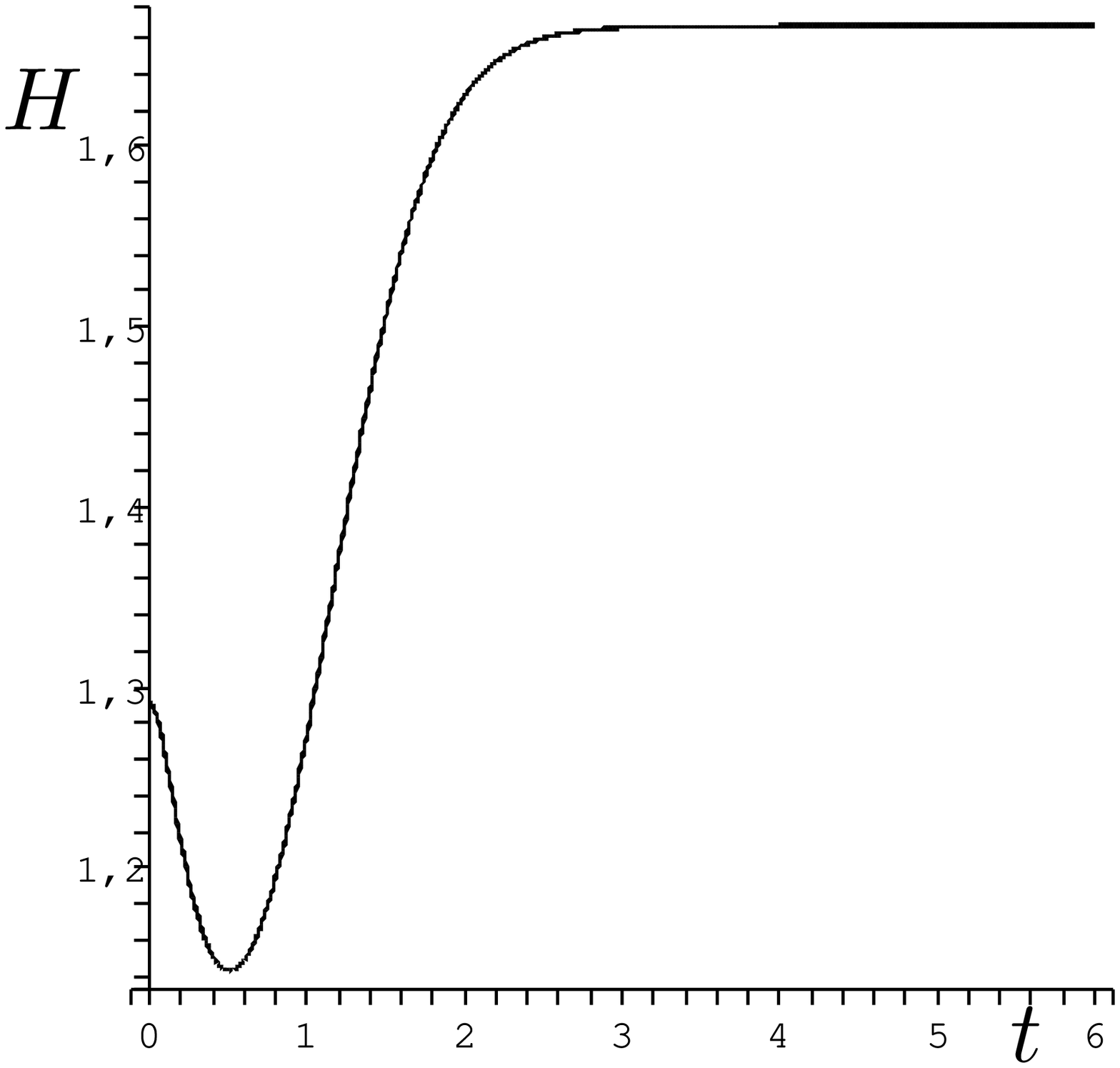} { }
\includegraphics[width=40mm]{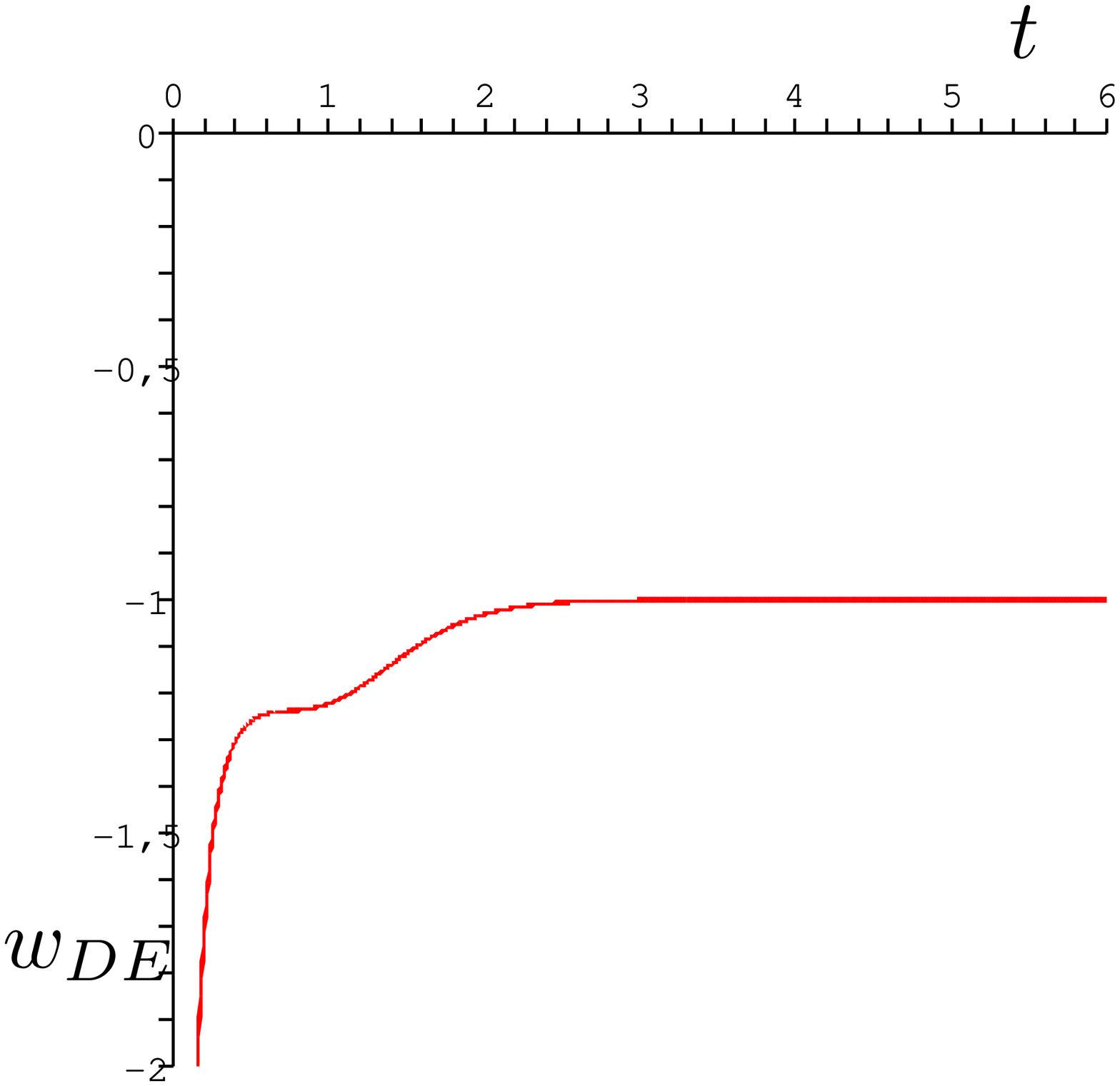}
\caption{The time evolution of the scalar field $\phi(t)$, the Hubble parameter $H(t)$ and the state
parameter $w_{DE}(t)$.
$\varrho_{M,0}=1$, $m_p^2=0.2$ and $\dot\phi_0=1$.} \label{plotsCDM1}
\end{figure}

\begin{figure}[h]
\centering
\includegraphics[width=40mm]{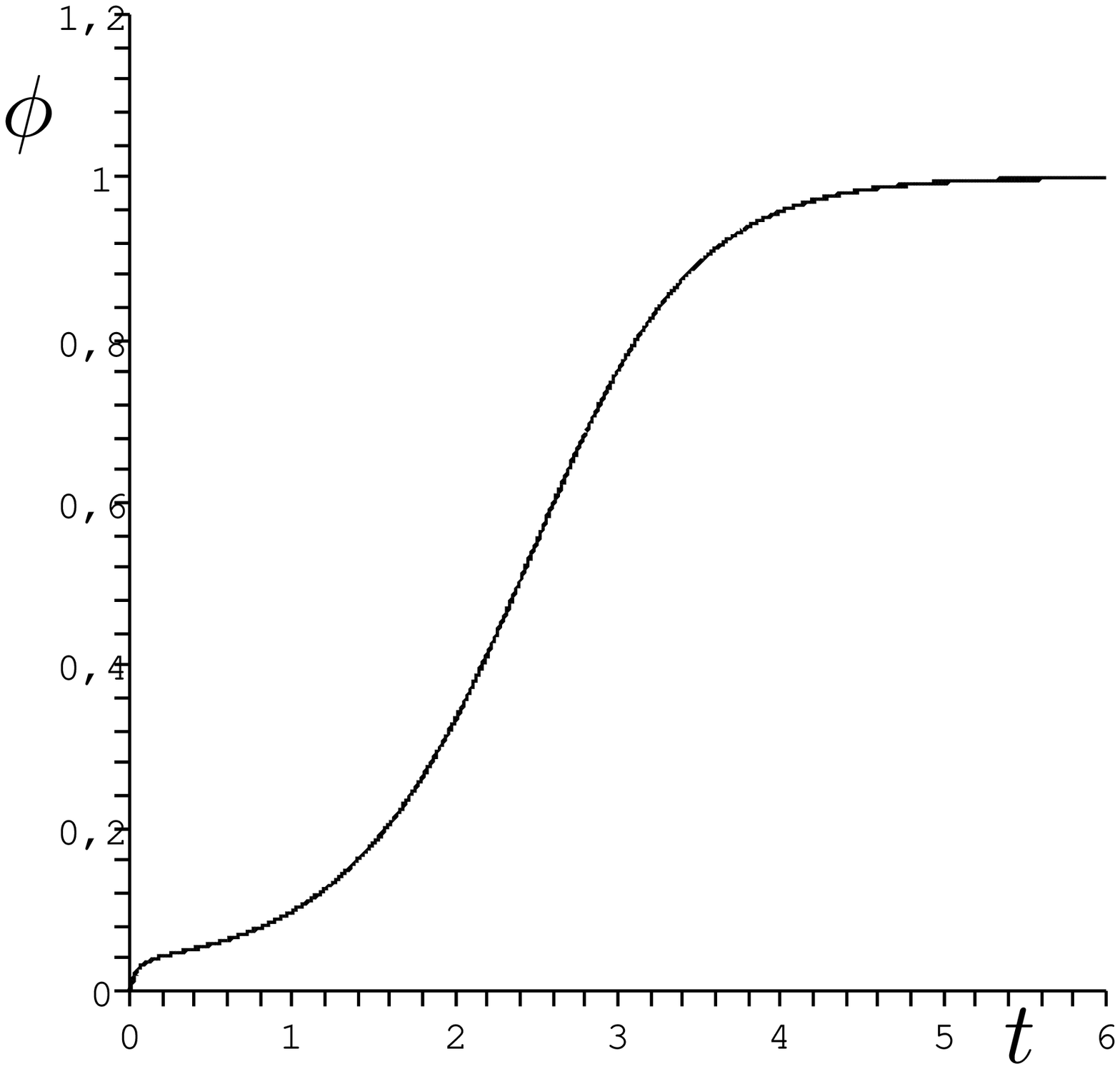} { }
\includegraphics[width=40mm]{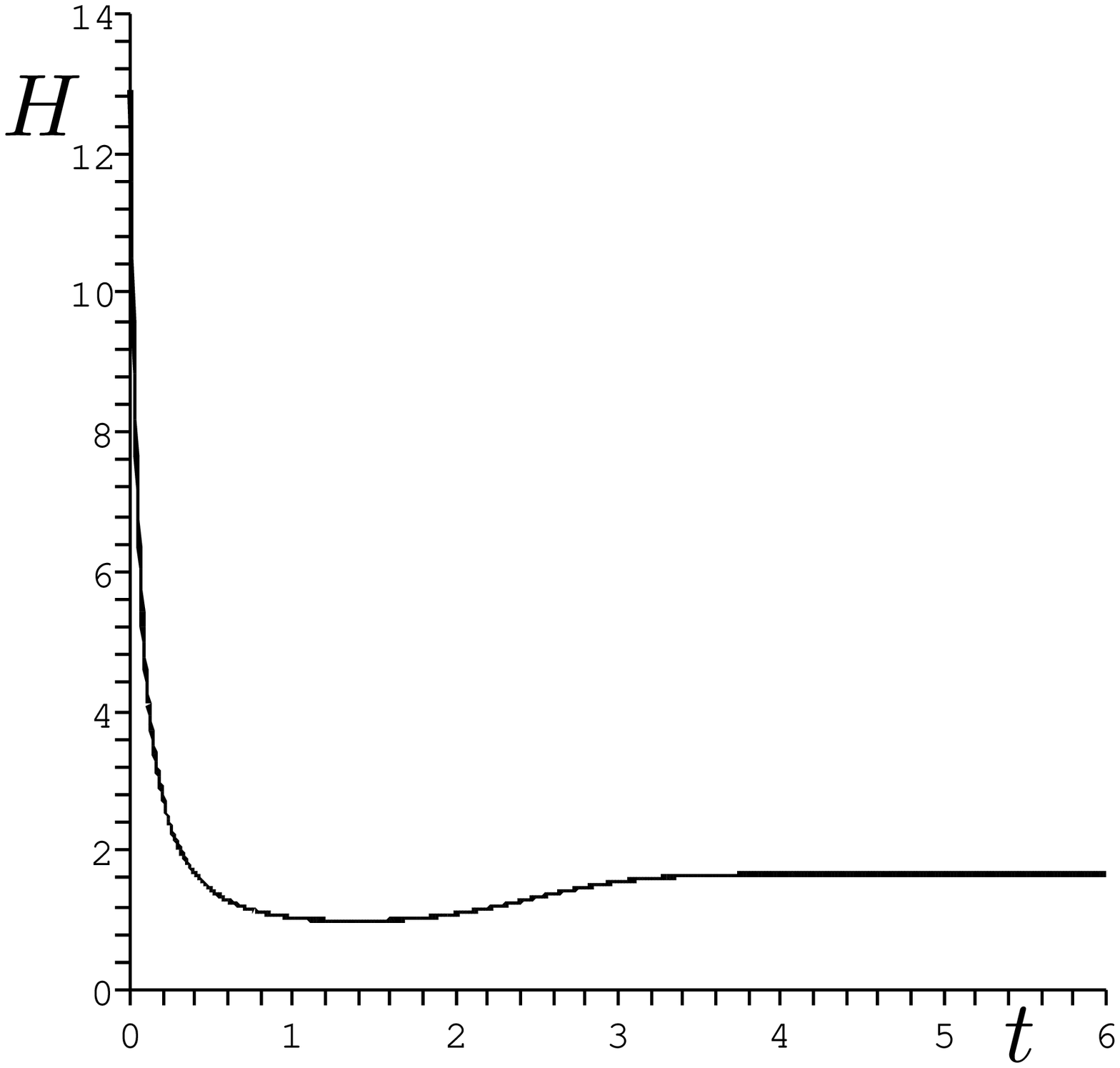} { }
\includegraphics[width=40mm]{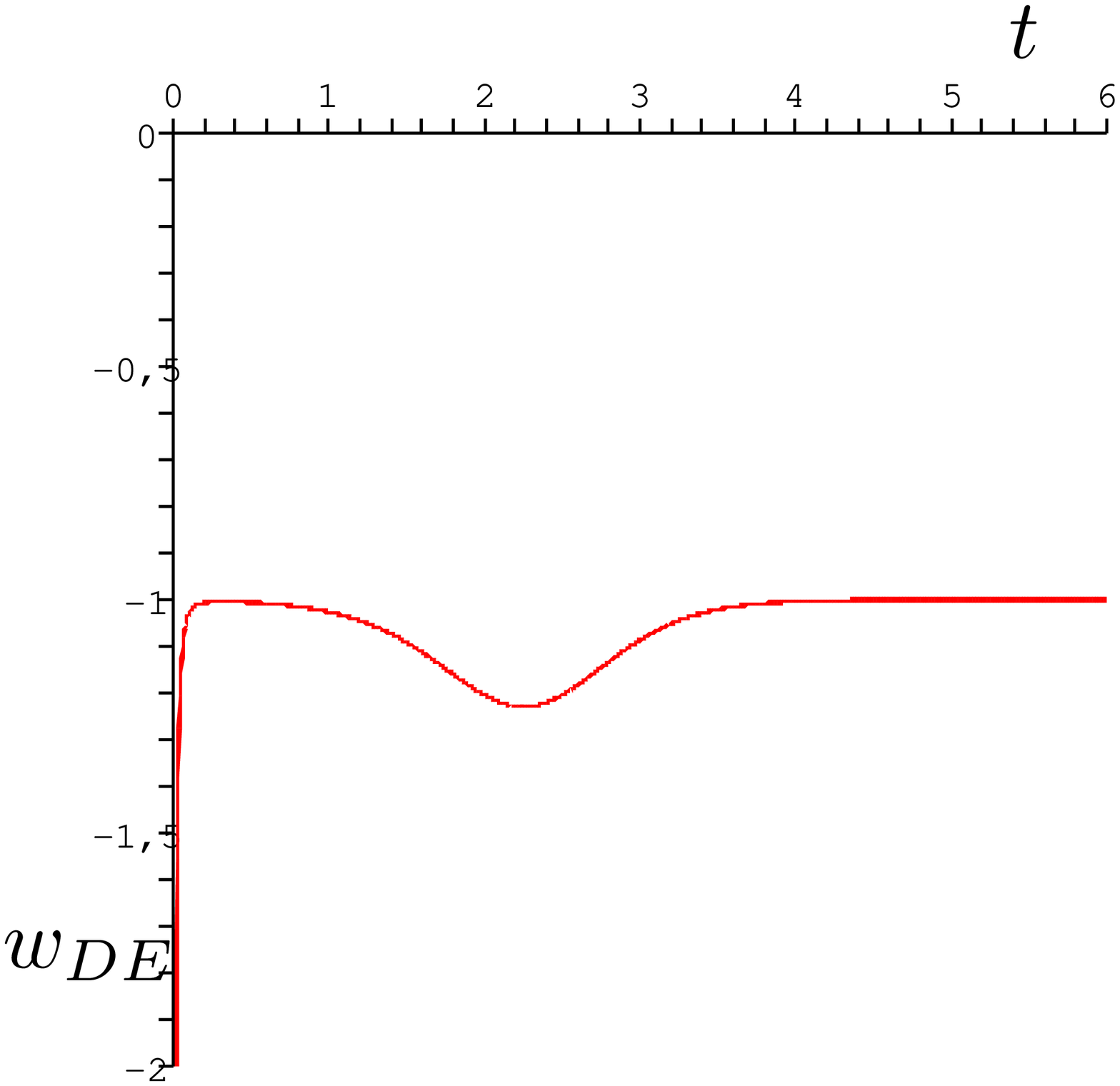}
\caption{The time evolution of the scalar field $\phi(t)$, the Hubble parameter $H(t)$ and the state
parameter $w_{DE}(t)$.
$\varrho_{M,0}=100$, $m_p^2=0.2$ and $\dot\phi_0=1$.}
\label{plotsCDM100}
\end{figure}

Due to equation (\ref{eom-cdm-expl}) initial data $\phi_0,~\dot\phi_0$ and
$H_0$ do fix an initial value of the CDM density. To have a given initial energy density
of the CDM we take $\phi_0$ and $\dot\phi_0$ and
find the corresponding value $H_0$. In particular, to have
$\varrho_{M,0}=1$, $\phi_0=0$, $\dot\phi_0=1$ for $m_p^2=0.2$ we must
take $H_0=\sqrt{5/3}\approx 1.29$. For this initial values numeric solutions
are presented graphically in Fig.\ref{plotsCDM1}.
In Fig.\ref{plotsCDM100} we present the same plots  for $\varrho_{M,0}=100$.
Corresponding $H_0=10\sqrt{5/3}$.
Comparing Figs.\ref{plotsCDM1} and \ref{plotsCDM100} with Fig.\ref{plots} one
can see that our solutions with and without the CDM are different only in the
beginning of the evolution where the CDM dominates (if exists). Note, that the
behavior of the Hubble parameter in the presence of the CDM is not monotonic and
the DE state parameter may be not monotonic as well.

\subsection{Numeric solutions. $\phi$-dependence.}

It turns out that values of the $w_{DE}$ as well as ratio
$\varrho_{CDM}/\varrho_{DE}$ which are observational cosmological
parameters \cite{Spergel,Tegmark,Seljak} can be found easier using equations
(\ref{fo-1}), (\ref{fo-2}) as functions of the e-folding number $n$. However,
it is more instructive to find a dependence on $\phi$ and not on $n$.

Let us recall that from an analysis of our phantom model without the CDM we
know that the scalar field interpolates between an unstable and a
nonperturbative vacua during infinite time similar to the non-BPS string
tachyon \cite{AJK,yar}.  In our notations nonperturbative vacuum corresponds
to $\phi=+1$. In the pure phantom model the evolution is described by
$\phi(t)=A\tanh(\omega t)$ function, where $A$ and $\omega$ can be rescaled to
1. This dependence is monotonic and this allows us to find  physical variables
as functions of $\phi$.

The situation is more complicated in the presence of the CDM. First, we do not
know an exact time dependence of the scalar field. Second, it is not evident
for arbitrary initial data and value of parameter $m_p^2$ that the scalar
field evolves monotonically. However, in the particular cases presented in
Figs. \ref{plotsCDM1} and \ref{plotsCDM100} our solutions $\phi(t)$ are
monotonic functions of time and moreover look like $\tanh(t)$ at large times.
Below we are interesting in solutions which approach the nonperturbative
vacuum during an infinite time. Thus, the point $\phi=1$ corresponds to an
infinite future. The $\phi(t)$ dependence can be found numerically to pass
from $\phi$ coordinate to the time.

\begin{figure}[h]
\centering
$~~~~~~~~\rho_{M,0}=0.01,~~~~~~~~~~~~~~~~~~~\rho_{M,0}=1~~~~~~~~~~~~~~~~~~~~~\rho_{M,0}=100$\\
$~$\\
$~$\\
\includegraphics[width=130mm]{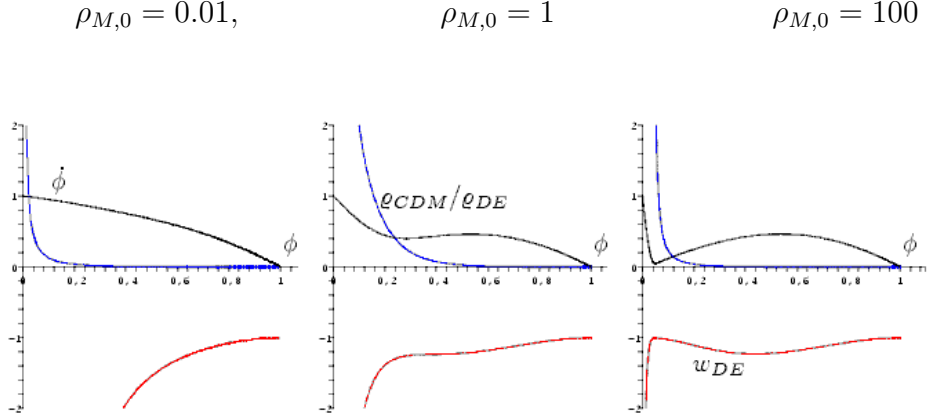}
\caption{
$\phi$-dependence of the velocity $\dot{\phi}$ (black line),
the state parameter $w_{DE}$ (red line) and the $\varrho_{CDM}/\varrho_{DE}$ ratio (blue line).
Initial velocity of the scalar field is equal to $1$ and
$m_p^2=0.2$.} \label{php1}
\end{figure}

\begin{figure}[h]
\centering
$m_p^2=0.2,~\rho_{M,0}=0.01~~~~~~m_p^2=0.2,~\rho_{M,0}=1~~~~~~m_p^2=0.2,~\rho_{M,0}=100$\\
$~$\\
$~$\\
\includegraphics[width=130mm]{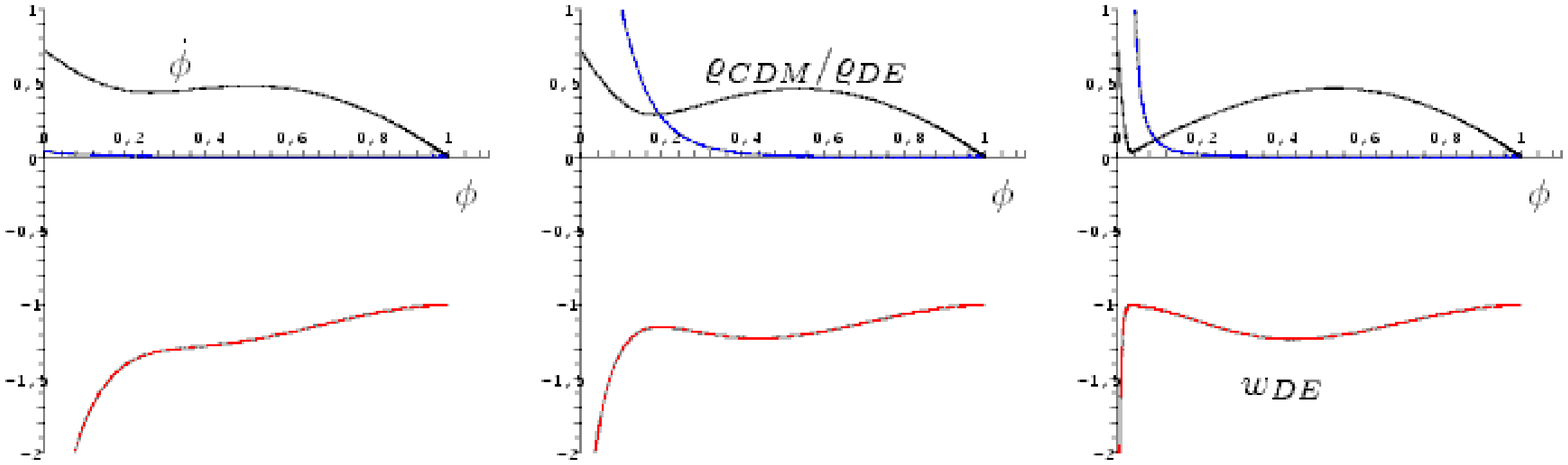} { }\\
$~$\\
$~$\\
$m_p^2=0.6,~\rho_{M,0}=0.01~~~~~~m_p^2=1,~\rho_{M,0}=0.01~~~~~~m_p^2=1,~\rho_{M,0}=1$\\
$~$\\
$~$\\
\includegraphics[width=40mm]{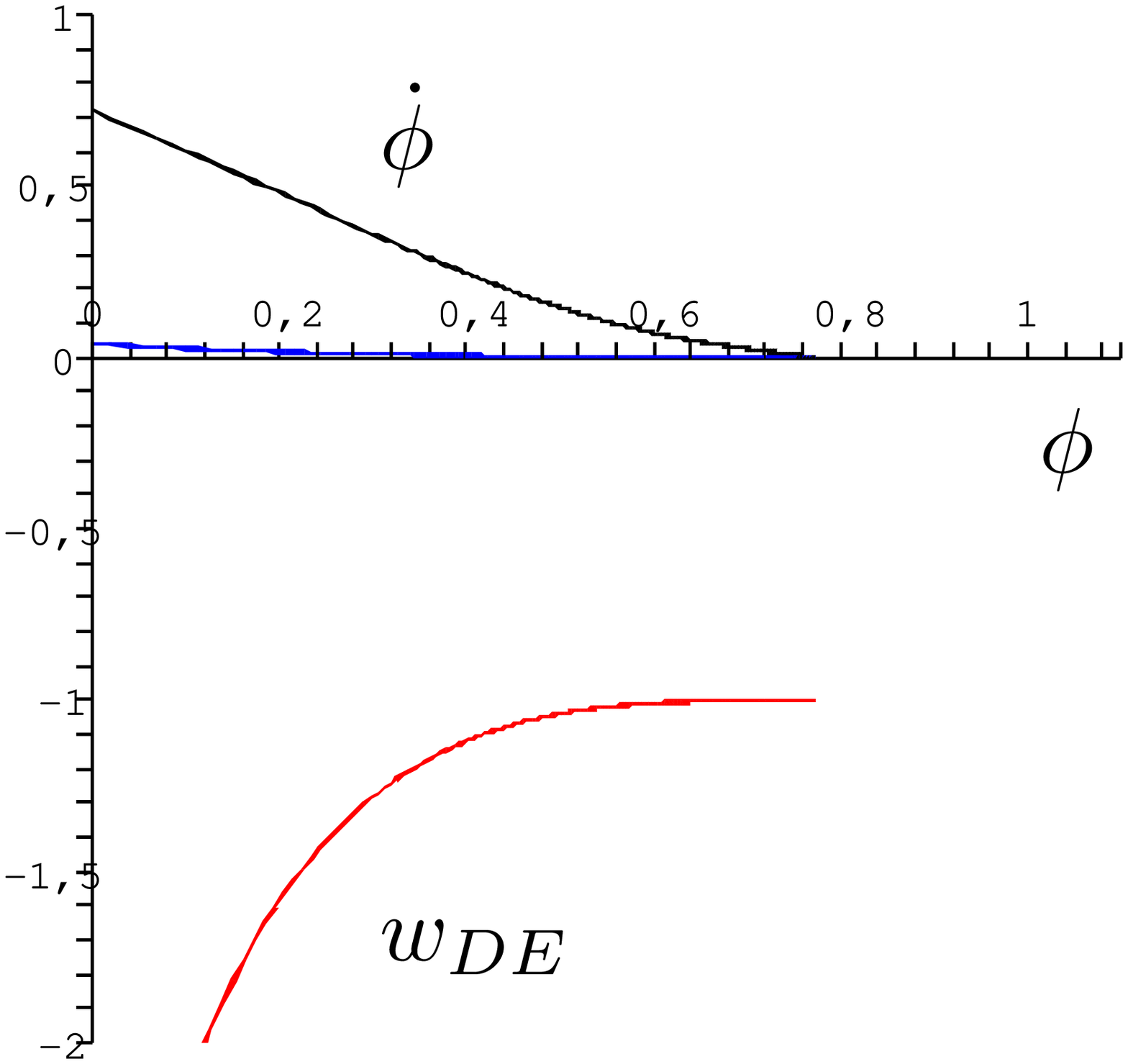} { }
\includegraphics[width=40mm]{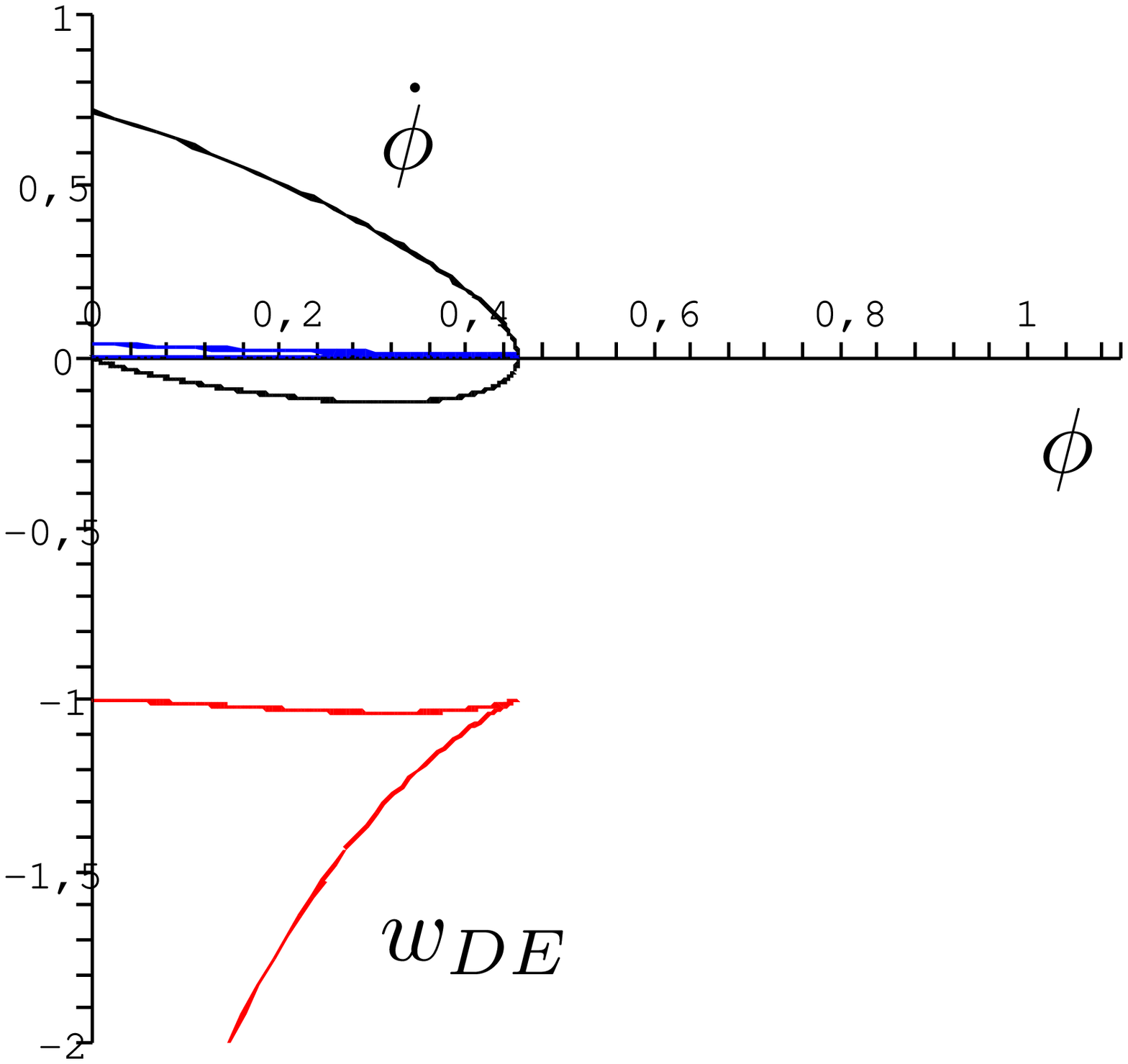} { }
\includegraphics[width=40mm]{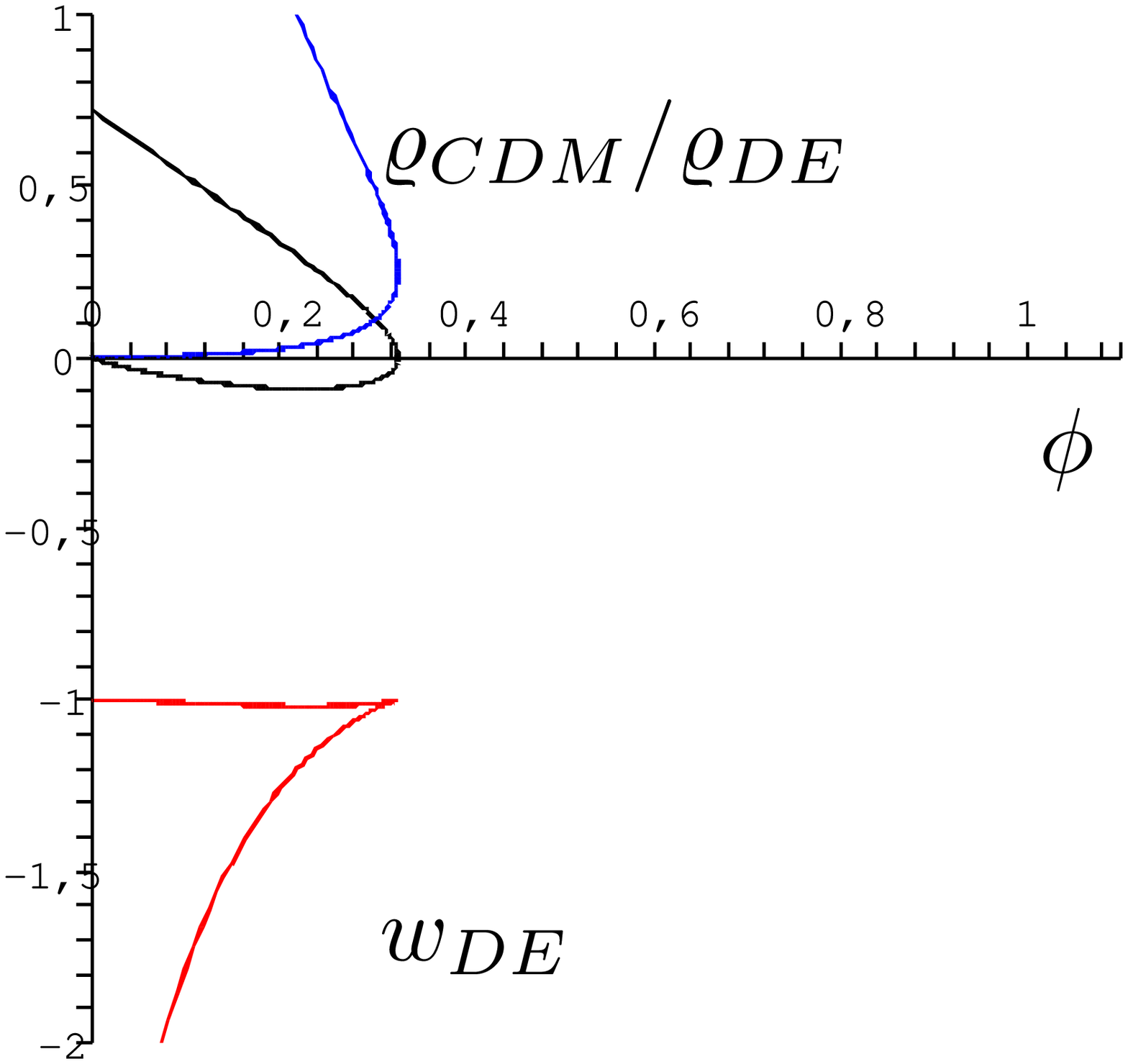} { }
\caption{
$\phi$-dependence of the velocity $\dot{\phi}$ (black line),
the state parameter $w_{DE}$ (red line) and the $\varrho_{CDM}/\varrho_{DE}$ ratio (blue line).
Initial velocity of the scalar field is equal to $0.72$.}
\label{php072}
\end{figure}

\begin{figure}[h]
\centering
$~~~~~~~~m_p^2=0.2,~~~~~~~~~~~~~~~~~~~m_p^2=0.6~~~~~~~~~~~~~~~~~~~m_p^2=1$\\
$~$\\
\includegraphics[width=40mm]{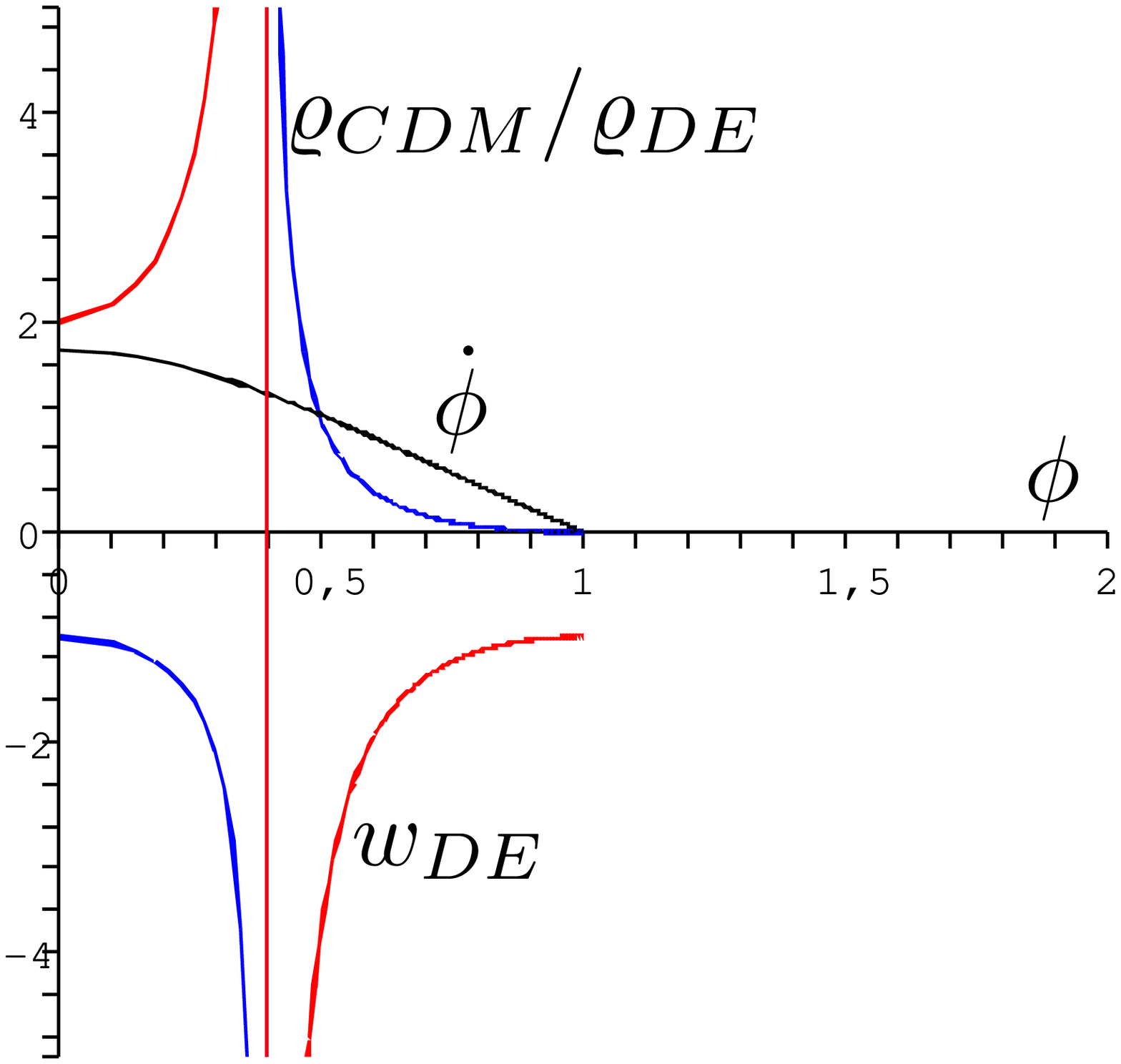}
\includegraphics[width=40mm]{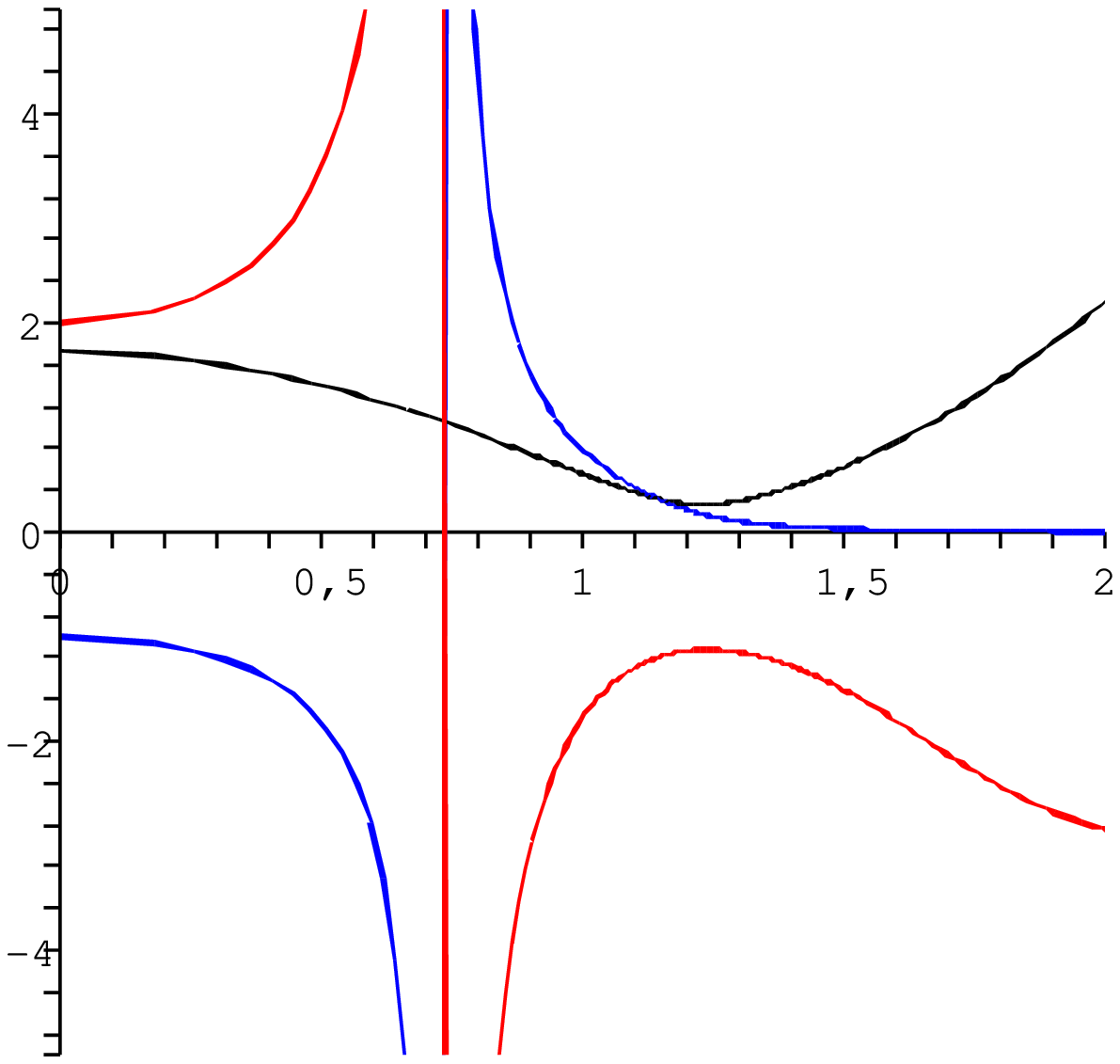}
\includegraphics[width=40mm]{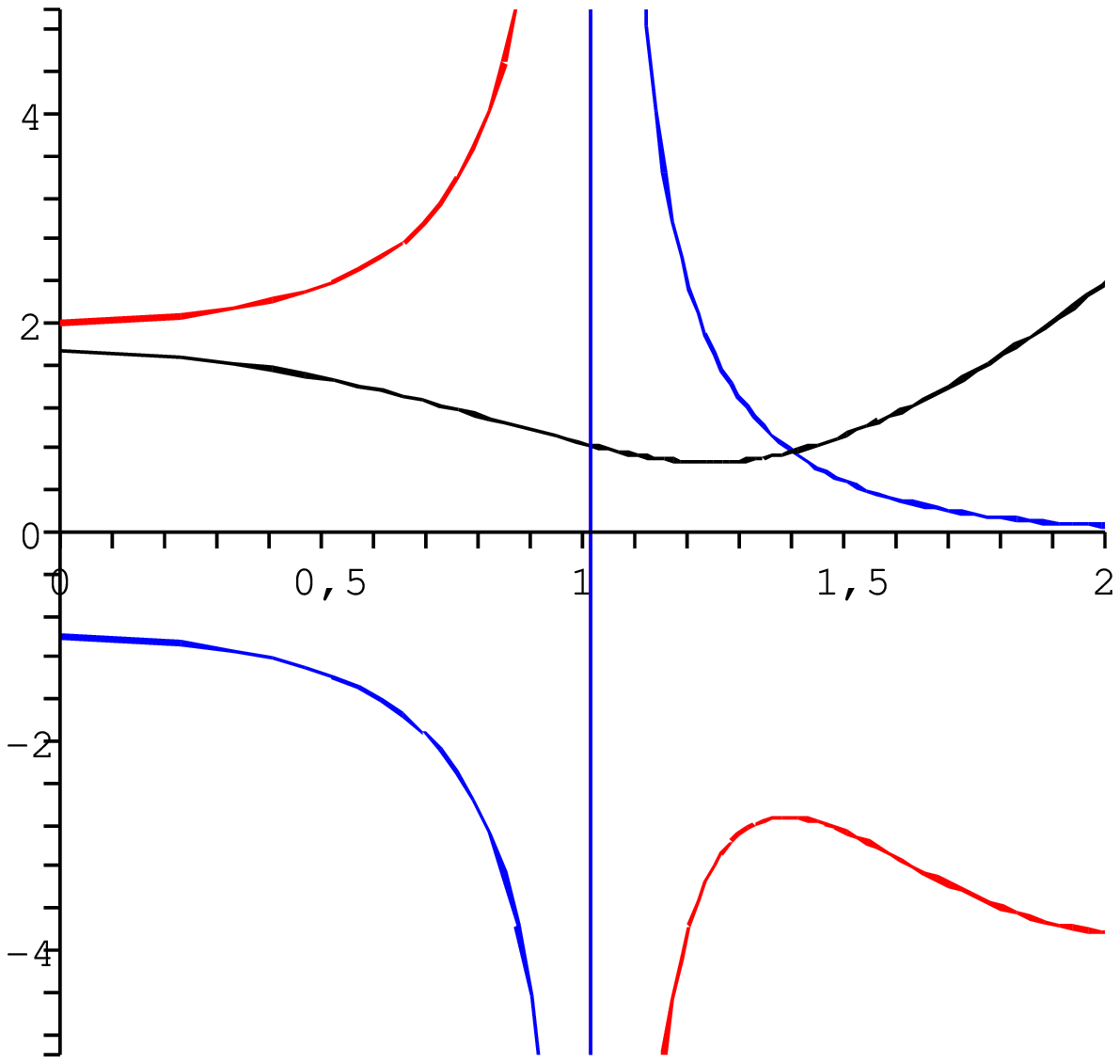}
\caption{$\phi$-dependence of the velocity $\dot{\phi}$ (black line),
the state parameter $w_{DE}$ (red line) and the $\varrho_{CDM}/\varrho_{DE}$ ratio (blue line).
Initial velocity of the scalar field is equal to its maximal possible
value.} \label{phpmax}
\end{figure}

In Figs.\ref{php1}--\ref{phpmax} we plot results of numeric solutions to
equations (\ref{fo-1}), (\ref{fo-2}) that allow us to find physical variables
such as $H$, $w_{DE}$, $\varrho_{CDM}/\varrho_{DE}$ as function of the field
$\phi$.
These sets of plots differ in an initial velocity of the scalar field. Note
that it follows from (\ref{eom-cdm-expl}) that there exists a maximal initial
velocity $\psi_{0m}$ for our phantom field. $\psi_{0m}$ depends on values of
$\phi_0$, $\varrho_{M,0}$ and $a_0$ and does not depend on $m^2_p$. In all
plots $a_0=1$ and $\phi_0=0$ . All plots have three curves: black ones are phase
curves, red ones are $w$-s and blue ones are $\varrho_{CDM}/\varrho_{DE}$ ratios. In
Fig. \ref{php1} the initial velocity is equal to $1$ (which is the same as for
the exact solution), $m_p^2=0.2$ and $\varrho_{M,0}$ is equal to $0.01$, $1$
and $100$ from left to right. Here we see that the scalar field reaches $+1$.
This indicates a stability of the system with respect to fluctuations of the
initial CDM energy density for small $m_p^2$. In Fig.\ref{php072} the initial
velocity is equal to $0.72$. The first row there corresponds to $m_p^2=0.2$
and $\varrho_{M,0}$ is equal to $0.01$, $1$ and $100$ from left to right. The
second row shows the behavior of the system with $m_p^2$ equal to $0.6$ and
$1$ and $\varrho_{M,0}$ equal to $0.01$ and with $m_p^2$ equal to $1$ and
$\varrho_{M,0}$ equal to $1$. One again sees from these plots that the scalar
field reaches $1$ for small values of $m_p^2$ in a wide range of an initial
CDM energy density.

This situation is broken for greater $m_p^2$ even for a small initial CDM
energy density and the field does not reach $1$. In Fig. \ref{phpmax}
$\varrho_{M,0}$ is taken to be $1$, the initial velocity is equal to its
maximal possible value $\psi_{0m}$ and  $m_p^2$ is equal to $0.2$, $0.6$ and
$1$ from left to right. Here we again observe a stability for small $m_p$ and
also find out that for large $m_p^2$ the scalar field goes beyond the point
$1$. Also for the maximal possible initial velocities $w_{DE}$ and
$\varrho_{CDM}/\varrho_{DE}$ functions have a discontinuity. One can
understand this qualitatively because the energy density for the DE has two
terms with opposite signs. Indeed, the scalar field is a phantom and its
kinetic energy is negative while the potential term is positive. Thus at some
point the energy density of the DE changes the sign and develops a
discontinuity in $w_{DE}$ and $\varrho_{CDM}/\varrho_{DE}$ ratio. Such a
behavior is rather undesirable from cosmological point of view, since the is
no observational data indicated singular behavior of cosmological parameters
and we do not consider corresponding plots further.

Hence, seeking for a situation where field $\phi$ approaches $1$ and there is
no cosmological singularities during this evolution we are left with the first
row in Fig. \ref{php1} and Fig. \ref{php072}. In this plots phase curves show
that field $\phi$ indeed depends monotonically on time because $\dot\phi$ is
always positive during the evolution. Looking for specified plots we draw the
reader's attention to the following interesting properties of our model.
First, $\varrho_{CDM}/\varrho_{DE}$ ratio dependence is monotonic and
experimentally measured value $1/3$ is close to the beginning of the
evolution. For example, in Fig.~\ref{php1} (left) this point corresponds to $w_{DE}\approx-1.02$ and
$\phi\approx 0.09$. Moreover, this is not a distinguished value and it follows from the
model that this ratio will decrease with time. Second, for $\rho _{M,0}$ large
enough $w_{DE}$ behaves non-monotonically.


\section{Discussion and Conclusion}
To summarize,
let us also note that we get an existence of a region of the initial
energy density of the CDM, for which $w_{DE}$ is not monotonic. Such a behavior is
interesting and very surprising. We see that for large initial energy densities of
the CDM $w_{DE}$ grows with time from minus infinity to approximately
$-1$, then goes down to a local minimum and after this grows again
asymptotically approaching $-1$. Note, that it has been proved in  \cite{vikman}
that under some (compare with \cite{andrianov}) conditions the phantom matter cannot cross the $w=-1$ barrier.
For small initial energy densities of the CDM
we cannot say that $w_{DE}$ has a local minimum but its rate of change becomes
slower and faster. We especially point out that the very first time stamp when
$w_{DE}$ approaches $-1$ (or its rate of change decreases essentially) is
approximately the same where $\varrho_{CDM}/\varrho_{DE}$ ratio is close to
$1/3$. It is worth to note that juxtaposing Figures
\ref{plotsCDM1}--\ref{plotsCDM100} and Figures \ref{php1}--\ref{phpmax}
 we see that the present ratio of the CDM and the DE
energy densities also corresponds to a local minimum of the Hubble parameter.

It is interesting to find an influence of the higher open string mass levels
as well as an influence of the closed string excitations on the obtained picture.
Even in the flat space-time the dynamics of a D-brane change drastically when
the closed string excitations are included \cite{Oh,LY}.

We get a stable behavior and smooth
cosmological parameters in the stringy inspired  model only in the case
when the dimensionless parameter $m^2_p$ is less than $0.5$. This
restricts the parameters of the original theory \cite{IA1} the
model considered in this paper comes from. Let us recall that $m^2_p$ is
related with the  reduced Planck mass, the string mass parameter $M_s^2$ and
the open string coupling constant $g_o^2$:
$$
m^2_p=\frac{g_o^2 M^2_P}{M^2_s}
$$
Therefore, to have an acceptable
cosmological solution we have to assume that $2 g_o^2 M^2_P<M^2_s$.
Here $M_P$ is the 4-dimensional Planck mass.
The effective string mass depends on the 10-dimensional mass
parameter $1/l_s$, where $l_s$ is a string length and the  compactification
volume $V_6$. Taking into
account that $M^2_s\sim V_6/l^8_s$ and using the usual requirement that
$l_s\sim 1/M_P$ we get
$$
m^2_p=\frac{g_o^2 M^2_P}{M^2_s}\sim \frac{g_o^2 M^2_P l^8_s}{V_6} \sim
\frac{g_o^2  l^6_s}{V_6}< 0.5,
$$
that looks reasonable from a general point of
view.


\section*{Acknowledgements}

This work is supported in part by RFBR grant 05-01-00758, I.A. and A.K. are
supported in part by INTAS grant 03-51-6346 and by Russian Federation
President's grant NSh--2052.2003.1, S.V. is supported in part by Russian
Federation President's Grant NSh--1685.2003.2 and by the grant of the
scientific Program "Universities of Russia" 02.02.503.


\end{document}